\documentclass{emulateapj}
\usepackage{graphicx}
\usepackage{natbib}
\usepackage{multirow}
\usepackage{afterpage}
\def\Lsol {$\hbox{L}_\odot$}
\def\Msol {$\hbox{M}_\odot$}
\def\kms {$\rm km~s^{-1}$}
\def\klambda {k$\lambda$}

\slugcomment{submitted \today}

\shorttitle{Orion Hot Core}
\shortauthors{Wright \& Plambeck}

\begin{document}

\graphicspath{ {/o/wright/OriALMA/HCpaper/figures/} }

\title{ALMA Images of the Orion Hot Core at 349 GHz}

\author{ M.~C.~H.~Wright, R.~L.~Plambeck}
\affil{Radio Astronomy Laboratory, University of California, Berkeley, CA 94720, USA}
\email{wright@astro.berkeley.edu}

\begin{abstract}

We present ALMA images of the dust and molecular line emission in the Orion Hot
Core at 349~GHz. At  0.2\arcsec\ angular resolution the images reveal multiple
clumps in an arc $\sim$ 1\arcsec\ east of Orion Source~I, the protostar at the
center of the Kleinmann-Low Nebula, and another chain of peaks from IRc7
towards the southwest.  The molecular line images show narrow filamentary
structures at velocities $> 10$~\kms\ away from the heavily resolved ambient
cloud velocity $\sim 5$~\kms.  Many of these filaments trace the SiO outflow
from Source~I, and lie along the edges of the dust emission.  Molecular line
emission at excitation temperatures 300--2000 K, and velocities $> 10$~\kms\
from the  ambient cloud, suggest that the Hot Core may be heated in shocks by
the outflow from Source~I or from the BN/SrcI explosion.  The spectral line
observations also reveal a remarkable molecular ring, $\sim 2 ''$ south of
SrcI, with a diameter $\sim$ 600 AU. The ring is seen in high excitation
transitions of HC$_3$N, HCN v2=1, and SO$_2$.  An impact of ejecta 
from the BN/SrcI explosion with a dense dust clump could
result in the observed ring of shocked material.

\end{abstract}

\vspace{6pt} 


\section{INTRODUCTION}
\label{sec:intro} 

The Kleinmann-Low (KL) Nebula in Orion, at a distance of 415~pc
\citep{Menten2007,Kim2008}, is the closest region in which high mass stars (M
$>8$~\Msol) are forming. The two most massive stars in the region, the
Becklin-Neugebauer Object (BN) and radio Source I (hereafter, ``SrcI'') appear
to be recoiling from one another at $\sim$ 40~\kms, argued to be the result of
a dynamic interaction between them 500 years ago that also launched an explosive
outflow of  at least 8~\Msol\ of material with velocities of 10 to $>$ 100~\kms\
\citep{Gomez2008,Zapata2009,Goddi2011,Bally2011,Bally2017}.  SiO emission also
traces a lower velocity outflow from SrcI that extends $\sim$~1000 AU into the
surrounding medium along a NE-SW axis (\citealp{Plambeck2009}; \citealp{Matthews2010,Greenhill2013}), roughly
perpendicular to SrcI's proper motion. 

Observations at many wavelengths, from cm to IR, reveal a dense Hot Core $\sim$
10\arcsec\ in diameter adjacent to SrcI.
The Hot Core does not appear to be internally heated; no known protostars are
embedded within it. 
\citet{Goddi2011b} mapped the Hot Core in 7 NH$_3$ lines with energy levels up
to 1500~K, and suggest that it is heated by the  outflow from ScrI and by the
impact of SrcI's proper motion towards it.

In this paper, we present 349~GHz, 0.2\arcsec\ resolution ALMA images of the
dust and molecular line emission from the Hot Core, comparing these with CARMA
images of the 229~GHz dust emission and the 86~GHz SiO v=0 outflow
\citep{Plambeck2013,Plambeck2009}.  These data suggest that both shocks from the SrcI outflow
and ejecta from the BN/SrcI explosion play a role in heating the Hot Core.

\section{OBSERVATIONS AND DATA REDUCTION}
\label{sec:obs}

The ALMA observations at 349 GHz, made in 2014 July, were designed to image the
continuum emission from SrcI, but were obtained in spectral line mode so that
line free portions of the spectrum could be identified.  Details of the ALMA
observations are given in \citet{Plambeck2016} Table~1.  Spectral line data
were obtained with $3840 \times 0.488$~MHz channels in 2~polarizations in each
of 4 spectral windows.

The visibility data were calibrated using ALMA-supplied scripts and the CASA
software package.  The calibrated CASA measurement sets then were written out
in FITS format and imported into {\tt Miriad}.  All further processing was done
with {\tt Miriad}, which took about 10\% of the time required by CASA.

As described in \citet{Plambeck2016}, the data were then self calibrated using
continuum images generated from the line-free spectral channels.  The
antenna-based phase corrections were derived only from visibilities  with
projected baselines greater than 250~k$\lambda$ in order to avoid confusion
from extended emission.  These images were dominated by emission from SrcI at
the center.  Visibilities measured on shorter spacings sample low spatial
frequency Fourier components of the source brightness distribution, primarily
thermal dust emission from large scale structures like the Orion Hot Core.  

\section{CONTINUUM IMAGES}

The channel-averaged visibilities for the 4 separate ALMA spectral windows were
combined using multifrequency synthesis to avoid bandwidth smearing at the edge
of the images.  The image was deconvolved using the {\tt CLEAN} algorithm.
Uncertainties in the fluxes are dominated by the absolute calibration
uncertainties of $\pm~10$\% for Band 7, as given by the ALMA Cycle 1 Technical
Handbook.

We compare the ALMA 349~GHz continuum image with a CARMA 229~GHz image with
similar angular resolution.  The 229 GHz data were obtained in the CARMA A- and
B-configurations in 2009 and 2011.  
The 229~GHz calibration is described in \citet{Plambeck2013}.  
The uv-coverage of the 229 and 349 GHz data is shown in \citet{Plambeck2016}, Figure~2.

\begin{figure} \centering

\includegraphics[width=0.9\columnwidth, origin=c, clip, trim=5cm 3.6cm 5cm 0cm] {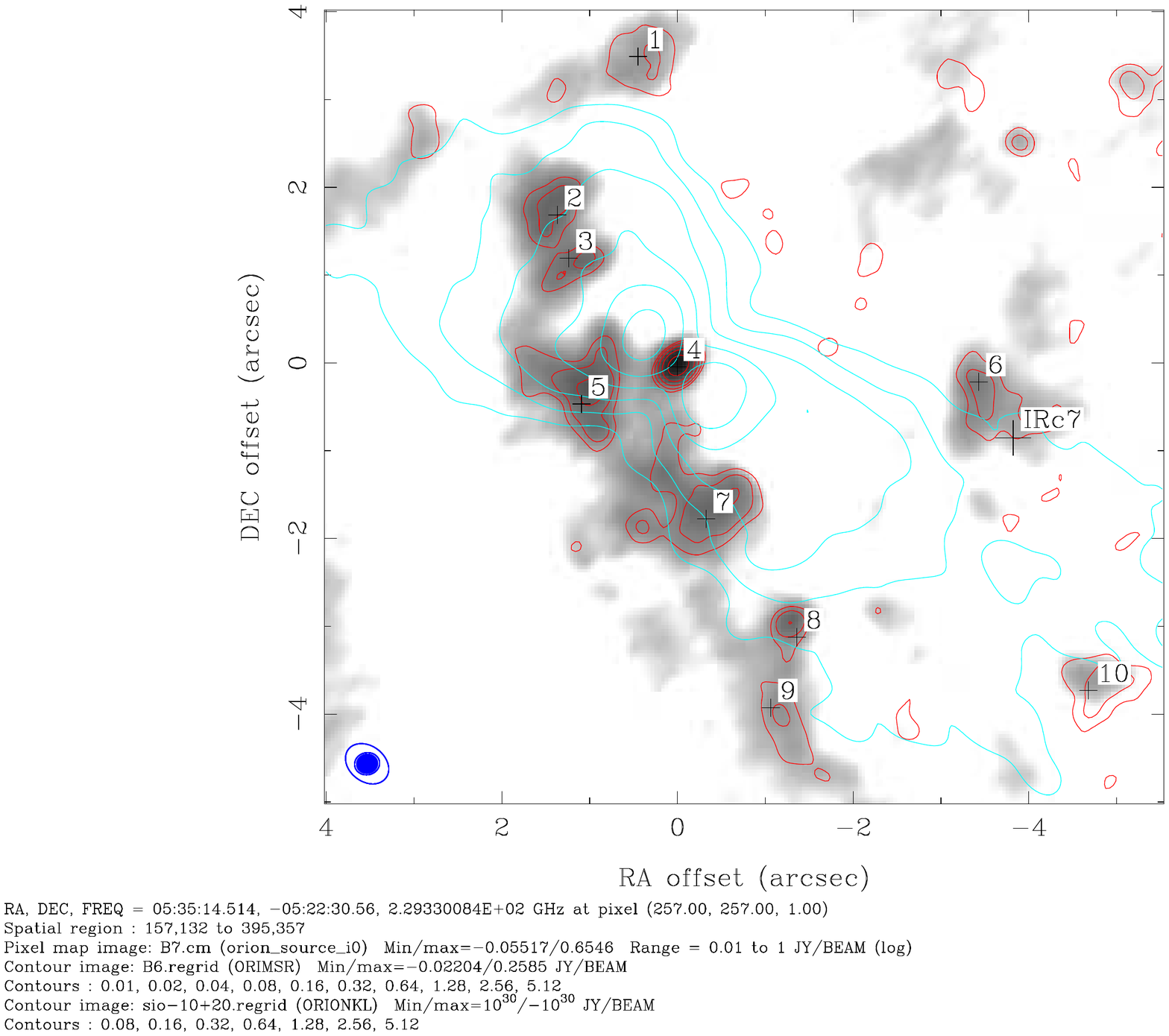} 

\caption{Comparison of 349 GHz~continuum imaged with ALMA (halftone) and the
229~GHz continuum (red contours) and 86~GHz SiO v=0 outflow (blue contours)
imaged with CARMA. The SiO emission is averaged over the velocity range -10 to
+20~\kms\ \citep{Plambeck2009}.  The halftone scale is logarithmic from 0.01 to
1 Jy/beam in order to show the low level emission that traces the boundaries of
the SiO outflow. Continuum peaks at 229 and 349~GHz agree well.  Compact
sources identified by \citet{Hirota2015} are marked by crosses; infrared source
IRc7 is associated with peak 6.  The contour levels are 1,2,4,8,16 $\times$ the
lowest levels, which are 0.01 and 0.08~Jy/beam at 229 and 86~GHz, respectively.
}

\label{fig:maps1} 
\end{figure}

Figure~\ref{fig:maps1} displays 349~GHz and 229~GHz continuum images 
that were generated from the self-calibrated data using all
baselines $>$40~\klambda, with {\tt robust=0.5} weighting.

There is good agreement between the 229 and 349 GHz emission, even though the
uv-coverage is quite different between these 2 radio telescope arrays.
Although our data do not fully sample the short spacings, the images at 349 and
229 GHz reveal a series of clumps of emission from the Hot Core in an arc
$\sim$ 1\arcsec\ east of SrcI, and another chain of peaks to the west, near the
infrared source IRc7.  These images are in good agreement with those presented by
\citet{Hirota2015}.  The compact sources identified by \citet{Hirota2015} are
marked by crosses in Figure~1.

Also shown in Figure 1 is the CARMA image of the 86 GHz SiO line in the
ground vibrational state \citep{Plambeck2009}, which traces the bipolar outflow
from SrcI.
The Hot Core continuum peaks lie along the
SE edge of the outflow, whilst IRc7 and several other continuum
peaks trace the W boundary of the  outflow.

\begin{deluxetable}{rlr}
\tablewidth{0.2\textwidth}
\tabletypesize{\small}
\tablecaption{Molecular Features}
\tablehead{
\colhead{freq}  & \colhead{molecule} & \colhead{E$_{\rm U}$}
}
\startdata
  340.7142 & SO &        81  \\
  340.8387 & $^{33}$SO & 87  \\
  355.0455 & SO$_2$ &   111   \\
  355.1865 & SO$_2$ &   180   \\
  356.0406 & SO$_2$ &   230   \\
  341.2755 & SO$_2$ &   369  \\
  342.7616 & SO$_2$ &   582   \\
  341.6740 & SO$_2$ &   679   \\
  341.4031 & SO$_2$ &   808  \\
  354.8000 & SO$_2$ v2=1 & 928 \\
  342.4359 & SO$_2$ v2=1 & 1041  \\
  343.9237 & SO$_2$ v2=1 & 1058 \\
  341.3233 & SO$_2$ &  1412  \\
  341.5594 & SO v=1 &  1686  \\
  354.6975 & HC$_3$N &   341 \\
  354.4604 & HCN v2=1 &  1067 \\
  356.2556 & HCN v2=1 &  1067  
\enddata
\label{tab:clumpology}
\end{deluxetable}

\begin{figure*} \centering
\includegraphics[width=0.95\textwidth, clip, trim=0.5cm 8.0cm 1cm 4.5cm ] {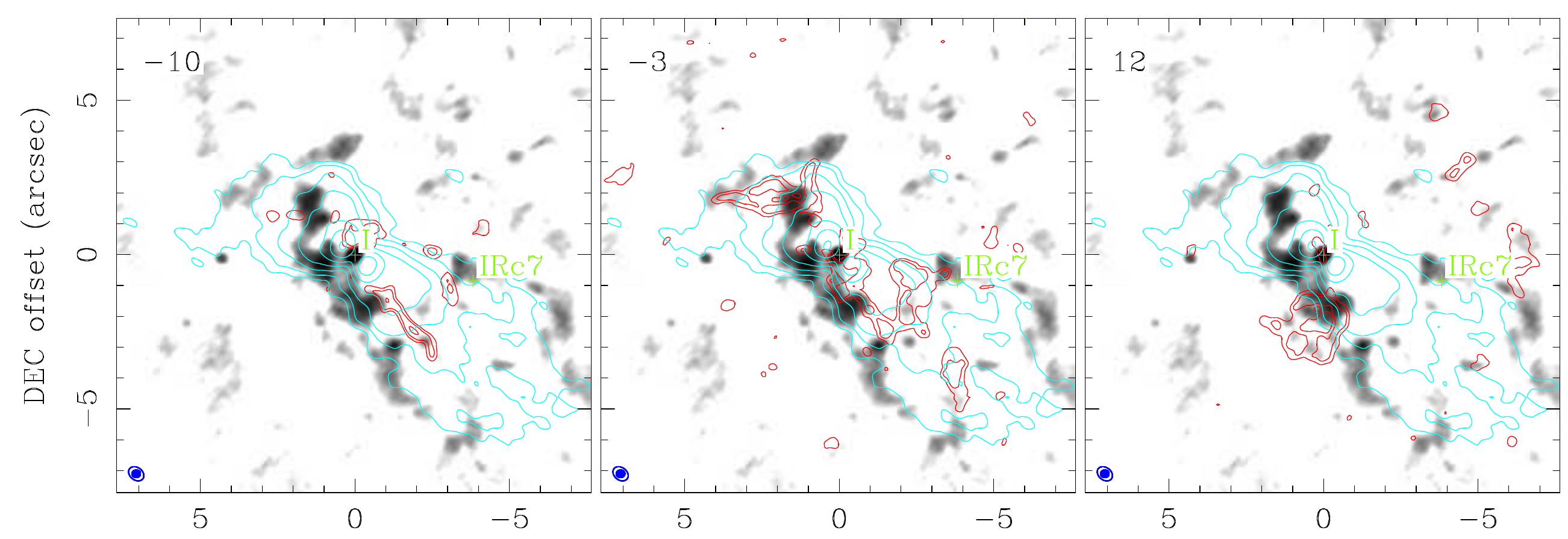} 
\includegraphics[width=0.95\textwidth, clip, trim=0.5cm 7.0cm 1cm 4.5cm ] {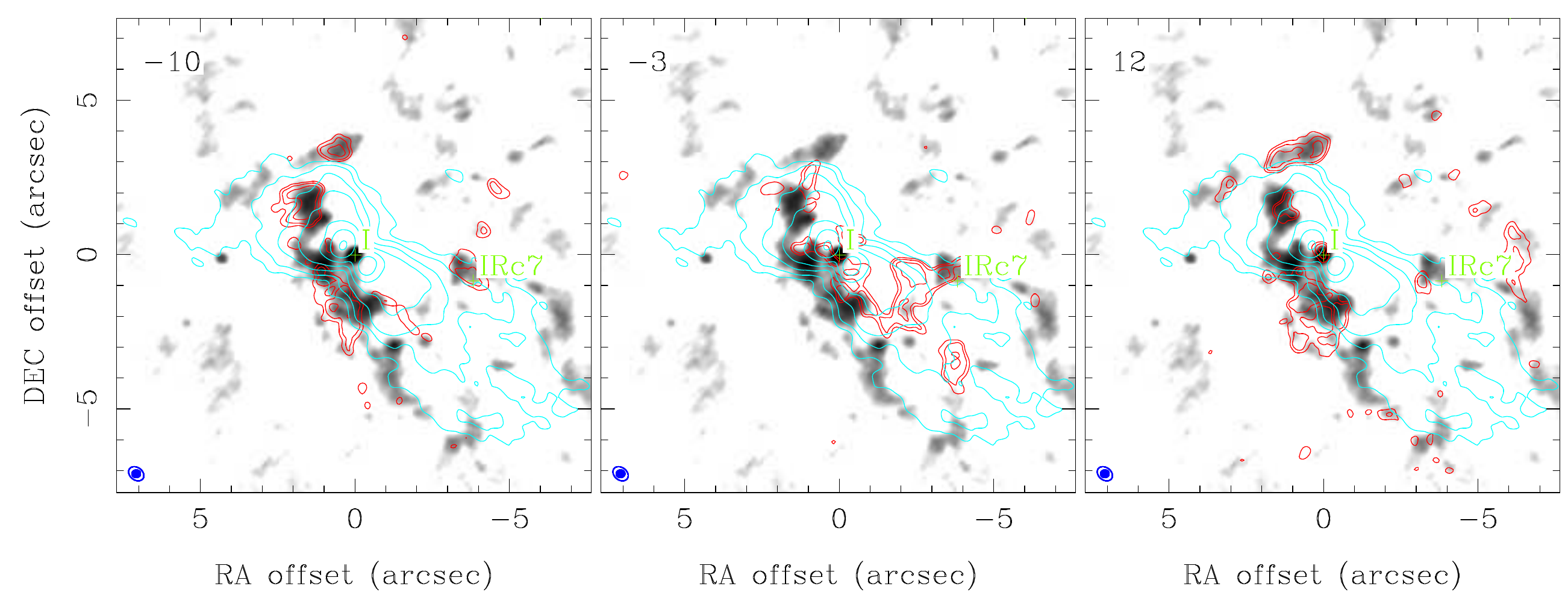} 

\caption{(top panels) Comparison of HC$_3$N emission with the 349 GHz dust
continuum and the 86~GHz SiO v=0 outflow from Orion SrcI.  The HC$_3$N line at
354.6975~GHz is shown by red contours, with levels 0.1, 0.2, 0.4, 0.8~~Jy/beam,
in 1~\kms\ wide channels centered at LSR velocities of -10, -3, and +12~\kms.
The 349~GHz dust continuum is indicated by the gray scale, logarithmic from
0.01 to 0.2 Jy/beam.  The SiO v=0 outflow is in blue contours, with levels
0.05, 0.1, 0.2, 0.4, $\ldots$, 3.2~Jy/beam. (bottom panels) Same, for the HCN
v=2 line at 356.2556~GHz, in red contours, again with levels 0.1, 0.2, 0.4,
0.8~Jy/beam.   In all panels the FWHM synthesized beam for the Band~7 spectral
lines is indicated by a filled blue ellipse, and for the SiO v=0 outflow by the
open ellipse.  }

\label{fig:maps2}
\vspace{12pt} 
\end{figure*}

\begin{figure} \centering
\includegraphics[width=0.9\columnwidth, clip, trim=4cm 3.6cm 4cm 0cm ] {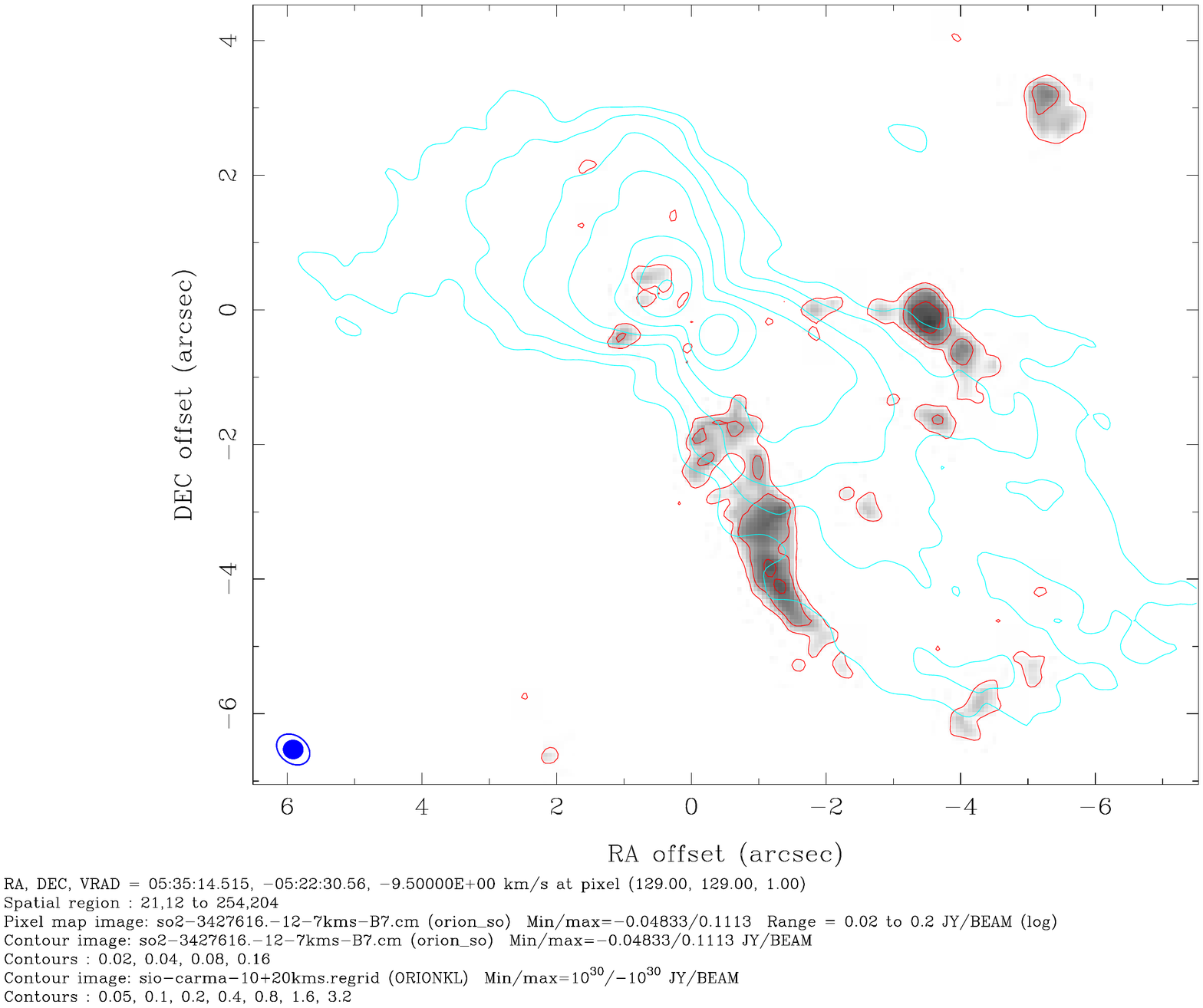} 

\caption{Molecular emission tracing the SiO outflow from SrcI.  SO$_2$ emission
at 342.7616 GHz in the velocity range -12 to -7 \kms\ is shown in red contour
levels 0.02 0.04 0.08 0.16 Jy/beam and log grey scale from 0.02 to 0.2 Jy/beam.
The SiO v=0 outflow is shown in blue contours as in Figure 1.  The synthesized
beam FWHM for the Band 7 spectral lines is shown as the filled blue ellipse,
and for the SiO v=0 outflow as the open ellipse.  }

\label{fig:maps3} 
\vspace{12pt} 
\end{figure}

\begin{figure*} \centering
\includegraphics[width=0.8\textwidth, clip, trim=0.5cm 3.5cm 1cm 1cm  ] {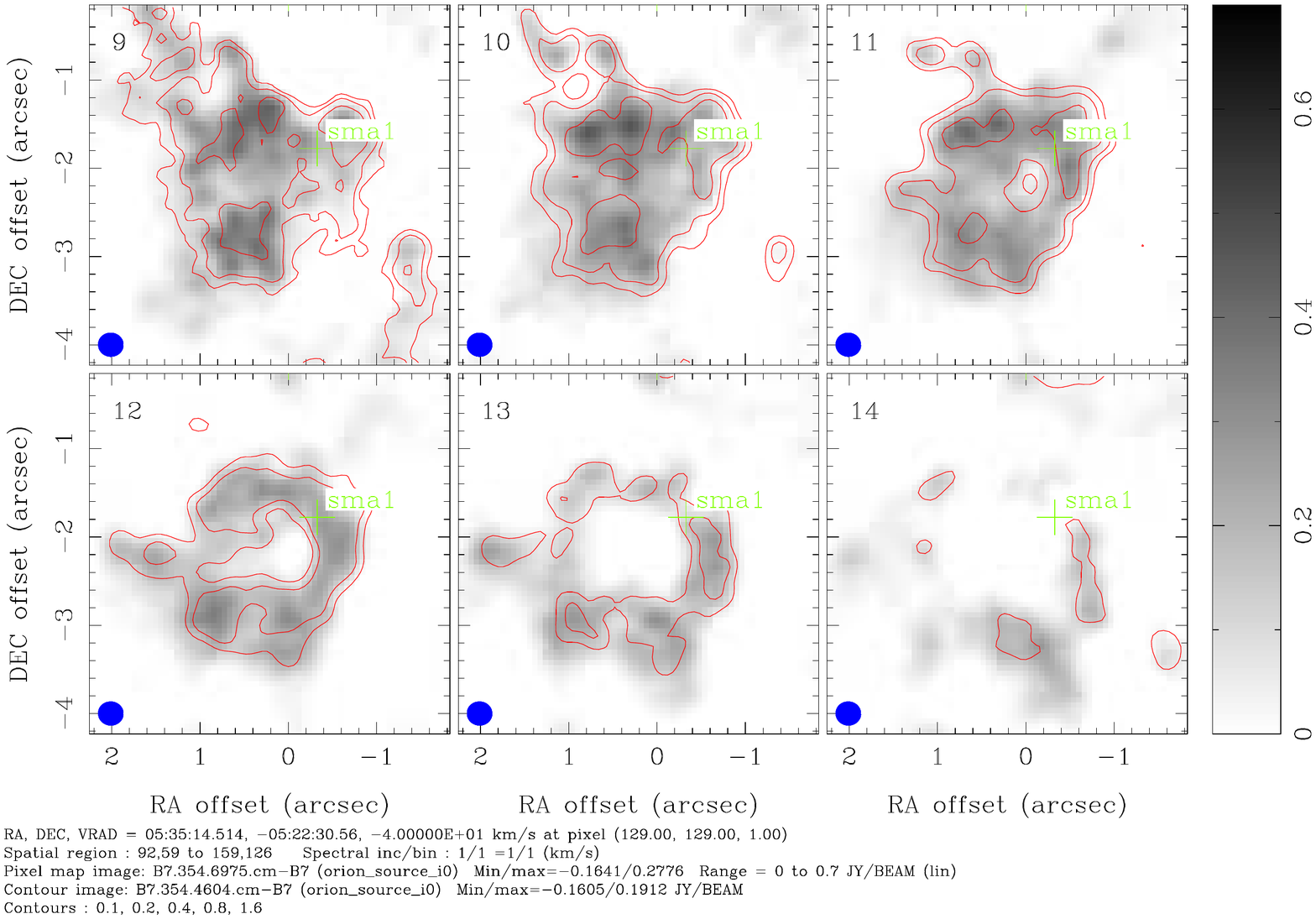} 

\caption{Molecular ring shown in HC$_3$N at 354.6975 GHz in grey scale, and HCN
v=2  at 354.4606 GHz in red contours in 1 \kms\ channels at velocities 9 to 14
\kms.  The grey scale is from 0 to 0.7 Jy/beam.  Band 7 continuum emission  has
been subtracted from the spectral lines.  The synthesized beam FWHM for the
Band 7 spectral lines is shown as the filled blue ellipse.  }

\label{fig:maps4} 
\vspace{12pt} 
\end{figure*}

\section{SPECTRAL LINES}
\label{sec:spectral}

Spectral line images were generated from the self-calibrated visibility data
using all baselines $>40$~k$\lambda$ and {\tt robust=0.5} weighting. We made
spectral line images for each of the 4 spectral windows; 4$\times$ 3840
channels in total.  A multitude of molecular lines are evident in the spectra.
\citet{Plambeck2016} show the spectra at the position of SrcI, and present
tables of the strongest spectral lines.  \mbox{Sulfur-} and \mbox{silicon-rich}
molecules such as SiO, SO, SO$_2$, SiS, and H$_2$S are particularly prominent
toward SrcI.  Examination of a wider field reveals a very confused picture with
deep sidelobes resulting from the poorly sampled large scale structures in
molecular line emission from the ambient molecular cloud. However, away from
the ambient cloud velocity many of the spectral lines show narrow filamentary
structures which appear to trace the  dust clumps and the edges of the SiO
outflow from SrcI.  

We made spectral line images at 1 \kms\ resolution in the velocity range -40 to
+40 \kms\ to image stuctures associated with the Orion Hot Core. Spectral
features which are {\it relatively} clear of confusing structure from other
spectral lines are listed in Table 1.  In Figure 2 we show emission in the
HC$_3$N and HCN v=2 transitions at 354.6975 and 356.2556 GHz in 1~\kms\
channels at  velocities -10, -3, and +12 \kms\ superposed on the SiO v=0
outflow. 

\section{DISCUSSION}
 
The images in Figures 1 and 2 provide important clues to the interaction of the
outflow from SrcI with the ambient molecular cloud, and to heating of the Hot
Core.  Dust emission seems to trace the periphery of the SiO outflow.  The Hot
Core peaks lie along an arc $\sim 1''$ east of SrcI where the SiO emission
drops off steeply (note the log contours).  Other ridges of dust emission are
found along the NE, SE, and SW boundaries of the outflow.  These observations
suggest that the dust emission originates from the walls of a cavity formed
by the outflow.

The  HC$_3$N and HCN v=2 molecular transitions shown in Figure 2 have upper
state energy levels of 341 and 1067~K. Often the peak molecular emission is
offset slightly from the dust clumps,
suggesting that the molecular gas may be heated by the interaction of the
outflow from SrcI with these clumps.  Many SO and SO$_2$ spectral lines with
E$_{\rm U}$ from 180 to 1058 K also trace the boundaries of the outflow. The
clearest example of this is shown in Figure 3 where we show the SO$_2$ line at
342.7616 GHz averaged over LSR velocities -12 to -7 \kms, overlayed on the the
SiO image.

\citet{Goddi2011b} imaged the Hot Core in NH$_3$ transitions with upper energy
levels 408 to 1456~K. Their observations show that the hottest gas lies in a
ridge $\sim 1''$ to the SE of SrcI. They suggest that the Hot Core may be
heated in C-type shocks driven by the outflow from SrcI, and by the proper
motion of SrcI to the SE at $\sim$ 12~\kms.  \citet{Peng2017} imaged the Hot
Core in 64 transitions of HC$_3$N and $^{13}$C isotopologues.  Their images of
vibrationally excited HC$_3$N lines also show that the excitation peaks lie
along the edge of the SiO outflow adjacent to the Hot Core. Both of these
studies argue for an external source of heating for the Hot Core.  Although
some of the continuum peaks 
could contain
embedded stars or protostars, spectral energy distributions for most of them
suggest masses of only 0.02 to 0.2 \Msol \citep{Hirota2015}.  Other observations in the
radio and infrared find no evidence for excitation peaks within the Hot Core
\citep{Menten1995, Greenhill2004b}.  Together, these observations suggest that
the outflow from SrcI may be heating the Hot Core and a shell of material tracing
the boundaries of the SiO outflow.

\subsection{Energetics}

SrcI is the dominant source of luminosity in KL.  At mm wavelengths SrcI
appears to be an optically thick dust disk with $\sim 0.1'' $ radius  and
temperature 500-700~K \citep{Plambeck2016,Hirota2016b}.  A 100~AU diameter disk
with temperature 700~K has a luminosity of $\sim 10^4$~\Lsol.
The Hot Core is a clumpy structure with diameter $\sim ~10''$ and a dust
temperature $\sim~100-150$~K, optically thick in the far infrared.  Modeling it
as a spherical black body with $\sim ~1000$ AU radius, its luminosity is (0.4-2) 
$\times 10^4$ \Lsol.
Thus, it appears that SrcI is capable of heating the Hot Core
to its observed temperature. 

Thermal excitation in 100-150~K gas does not, however, easily account
for the high excitation (E$_{\rm U} > 500$~K) HCN and SO$_2$ transitions in
Figures 2 and 3, or the HC$_3$N and NH$_3$ lines imaged by \citet{Peng2017} and
\citet{Goddi2011b}. \citet{Goddi2011b} suggest that the Hot Core is
heated by shocks from the impact of the outflow from ScrI and its proper motion
towards dense clumps in the Hot Core.  
The prominent abundance of SO and SO$_2$ transitions also is indicative of
shock chemistry. Models of C-type shocks with velocites 5-40~\kms\ can enhance
SO and SO$_2$ abundance by 2 orders of magnitude in the shock and extensively
in the post shocked gas \citep{Pineau1993}.

We now consider whether the mechanical luminosity of the SrcI outflow
is sufficient to account for these shock-excited transitions.  
\citet{Wright1995} estimate the mass outflow from the  J=2-1, v=0 SiO
outflow $\sim 10^{-5}$ \Msol~y$^{-1}$.  A similar value ($5~10^{-6}$) was
obtained for the SiO J=1-0 v=1 masers by \citet{Greenhill2013}. With a
velocity $\sim$ 18~\kms, the outflow luminosity is of order 1~\Lsol.  
Assuming isotropic radiation and a velocity width of $\sim$~20~\kms, the
luminosity of the high excitation HCN and SO$_2$ lines in Figures 2 and 3 is $\sim$ 1-10 $\times
10^{-5}$~\Lsol.  Hundreds, or perhaps thousands, of such lines are 
emitted by the Hot Core.  Although it appears that the outflow does have
sufficient mechanical luminosity to power these lines, a rather high
conversion efficiency would be required.


\subsection{Co-moving gas with SrcI}

The explosive outflow associated with the dynamic interaction of BN
and SrcI may also play a role in heating the Hot Core.
Several studies suggest that the  BN/KL outflow traced by fast moving CO and
H$_2$ bullets is associated with a wide angle outflow of  5--10 \Msol\ of gas
with velocities $\sim 20~$\kms\ and may have a common origin in a dynamic
interaction which ejected BN  to the NW and SrcI to the SE  $\sim 500$ years
ago \citep{Gomez2008,Zapata2009,Goddi2011,Bally2011,Bally2017}.  The kinetic
energy of the BN/KL outflow is estimated as $\sim 4 \times 10^{46}$
\citep{Snell1984} to $4 \times 10^{47}$ erg \citep{Kwan1976}.  The $\sim
20~$\kms\ outflow and the fast moving bullets appear to be blocked by the Hot
Core \citep{Chernin1996,Zapata2011a}.


The impact of this outflow on the Hot Core may  also be responsible 
for shock heating of high excitation molecules \citep{Zapata2011a,Goddi2011b}.
\citet{Zapata2011a} note that the high excitation lines HC$_3$N 37-36 v7=1 and
CH$_3$OH $7_{4,3}-6_{4-3}~v_t = 2$ appear to form a shell around the Hot Core
peaks, as is also observed for HC$_3$N and HCN in our ALMA observations.
It may be possible to identify the origin of the molecular material:
apparently oxygen-rich material (SiO, H$_2$O, OH, SO, SO$_2$), from the
SrcI outflow, and NH$_3$, HC$_3$N, HCN, CH$_3$CN, etc. from the Hot Core.

If the motion of SrcI is accompanied by co-moving gas, this helps to explain
why the SiO outflow from SrcI is not swept back by its motion though the
ambient gas. 
Magnetic support has been invoked to explain the collimation of the SrcI
outflow into a straight bipolar flow \citep[e.g.,][]{Greenhill2013,Vaidya2013},
and millimeter wavelength polarization observations \citep{Tsuboi1996,Plambeck2003} suggest a polarization parallel to the SrcI outflow, but the field strength is as yet undetermined, and it is not clear if magnetic fields control the direction of the
outflow, or vice versa.

Any original disk around SrcI was stripped off in the BN/SrcI encounter \citep{Bally2017}.
Co-moving gas also helps re-formation of a disk around SrcI through
Bondi-Hoyle, B-H, accretion.   The B-H radius $R_{BH} \sim GM/v^2$. The
accretion rate $\dot{M}_{BH} = \pi \rho R^2 v$ so $\dot{M} \sim  1/ v^3$  and
B-H accretion could re-form a  disk around SrcI if the relative velocity of
SrcI w.r.t. the ambient gas is only a few \kms. For example, with a relative
velocity of 2~\kms\ through an ambient density $n(\rm H_2) \sim 10^6$~cm$^{-3}$, SrcI
could accrete a disk mass  $\sim 0.07$~\Msol, consistent with an estimated 0.02 -
0.2~\Msol\ for the dusty disk model in \citep{Plambeck2016}.  

\subsection{Molecular Ring}

Figure 4 shows a molecular ring, $\sim 2 ''$ south of SrcI and  centered at  RA
= 05:35:14.535 DEC= -05:22:32.7 $\pm 0.1''$  in HC$_3$N at 354.6975 GHz  and HCN
v=2 at 354.4606~GHz in 1 \kms\ channels at velocities  from 9 to 14 \kms. The
ring presents an almost complete shell with a diameter $\sim  2.6 \times 1.9''$
in position angle -30 to -50 $ \deg$. The shell width decreases from 10 to 14
\kms with a velocity gradient across the minor axis in PA $\sim 45 \deg$.  The
ring can also be seen in the SO and SO$_2$ lines listed in Table 1 with E$_U$
from 180 to 678 K. It is not apparent in the lower excitation SO$_2$ lines
where the emission may be optically thick, or in the higher excitation lines
with E$_U > $800~K.  The high energy levels suggest shock excitation, perhaps
from high velocity ejecta from the BN/SrcI explosion.  A rotating, expanding
shell model is possible. There is no evidence for the red-shifted cap of the
ring at velocities $>$ 14~\kms, but this could have been dispersed by shear
motions of the outflow from SrcI. 
 An expansion velocity $\sim$ 2 \kms\ with a
ring radius $\sim$ 300 AU  gives a dynamical age $\sim$ 700 yr; a rotation velocity
$\sim$ 2 \kms gives an equilibrium mass $\sim$ 0.2 \Msol.
If the ring resulted from an impact on a dust clump with ejecta from the
BN/SrcI explosion $\sim$ 500 years ago then the ring has been decelerated by
swept up material.
C-type shocks with velocities as low as
10~\kms\ can completely remove the grain mantles and produce gas kinetic
temperatures to 500 K \citep{Flower1994}. 
 Some of the compact features in the Hot Core identified by \citet{Hirota2015}  might be ejecta from the BN/SrcI explosion.
We did not detect any continuum
emission at the center of the ring in any of the observations reported here.
The dust source SMA1, Hirota source number 7, is at the edge of the ring as
shown in Figure 4.

\section{SUMMARY}
\label{sec:summary}

The Orion Hot Core was imaged with ALMA at 349 GHz with $\sim$0\farcs2 angular
resolution.  Our observations offer support for a model where the Hot Core is
heated by shocks produced by the outflow from SrcI and from the BN/KL explosive
event.  A measurement of the proper motion of the most compact features in the
Hot Core clumps w.r.t. SrcI, if technically feasible, could reveal whether the
Hot Core itself is party to the expansion resulting from the BN/SrcI encounter.

\acknowledgments

We would like to thank an anonymous referee for helpful comments which
have improved this paper.
This paper makes use of the following ALMA data:
ADS/JAO.ALMA\#2012.1.00123.S. ALMA is a partnership of ESO (representing
its member states), NSF (USA) and NINS (Japan), together with NRC
(Canada) and NSC and ASIAA (Taiwan), in cooperation with the Republic of
 Chile. The Joint ALMA Observatory is operated by ESO, AUI/NRAO and NAOJ.

The National Radio Astronomy Observatory is a facility of the National
Science Foundation operated under cooperative agreement by Associated
Universities, Inc."

Support for CARMA construction was derived from the states of California,
Illinois, and Maryland, the James S. McDonnell Foundation, the Gordon and Betty
Moore Foundation, the Kenneth T. and Eileen L. Norris Foundation, the
University of Chicago, the Associates of the California Institute of
Technology, and the National Science Foundation.

{\it Facilities:} \facility{CARMA}, \facility{ALMA}.
\bibliographystyle{apj}
\bibliography{hc}

\begin{thebibliography}{}
\expandafter\ifx\csname natexlab\endcsname\relax\def\natexlab#1{#1}\fi

\bibitem[{{Bally} {et~al.}(2011){Bally}, {Cunningham}, {Moeckel}, {Burton},
  {Smith}, {Frank}, \& {Nordlund}}]{Bally2011}
{Bally}, J., {Cunningham}, N.~J., {Moeckel}, N., {et~al.} 2011, \apj, 727, 113

\bibitem[{{Bally} {et~al.}(2017){Bally}, {Ginsburg}, {Arce}, {Eisner},
  {Youngblood}, {Zapata}, \& {Zinnecker}}]{Bally2017}
{Bally}, J., {Ginsburg}, A., {Arce}, H., {et~al.} 2017, \apj, 837, 60

\bibitem[{{Chernin} \& {Wright}(1996)}]{Chernin1996}
{Chernin}, L.~M., \& {Wright}, M.~C.~H. 1996, \apj, 467, 676

\bibitem[{{Flower} \& {Pineau des Forets}(1994)}]{Flower1994}
{Flower}, D.~R., \& {Pineau des Forets}, G. 1994, \mnras, 268, 724

\bibitem[{{Goddi} {et~al.}(2011{\natexlab{a}}){Goddi}, {Greenhill},
  {Humphreys}, {Chandler}, \& {Matthews}}]{Goddi2011b}
{Goddi}, C., {Greenhill}, L.~J., {Humphreys}, E.~M.~L., {Chandler}, C.~J., \&
  {Matthews}, L.~D. 2011{\natexlab{a}}, \apjl, 739, L13

\bibitem[{{Goddi} {et~al.}(2011{\natexlab{b}}){Goddi}, {Humphreys},
  {Greenhill}, {Chandler}, \& {Matthews}}]{Goddi2011}
{Goddi}, C., {Humphreys}, E.~M.~L., {Greenhill}, L.~J., {Chandler}, C.~J., \&
  {Matthews}, L.~D. 2011{\natexlab{b}}, \apj, 728, 15

\bibitem[{{G{\'o}mez} {et~al.}(2008){G{\'o}mez}, {Rodr{\'{\i}}guez}, {Loinard},
  {Lizano}, {Allen}, {Poveda}, \& {Menten}}]{Gomez2008}
{G{\'o}mez}, L., {Rodr{\'{\i}}guez}, L.~F., {Loinard}, L., {et~al.} 2008, \apj,
  685, 333

\bibitem[{{Greenhill} {et~al.}(2004){Greenhill}, {Gezari}, {Danchi}, {Najita},
  {Monnier}, \& {Tuthill}}]{Greenhill2004b}
{Greenhill}, L.~J., {Gezari}, D.~Y., {Danchi}, W.~C., {et~al.} 2004, \apjl,
  605, L57

\bibitem[{{Greenhill} {et~al.}(2013){Greenhill}, {Goddi}, {Chandler},
  {Matthews}, \& {Humphreys}}]{Greenhill2013}
{Greenhill}, L.~J., {Goddi}, C., {Chandler}, C.~J., {Matthews}, L.~D., \&
  {Humphreys}, E.~M.~L. 2013, \apjl, 770, L32

\bibitem[{{Hirota} {et~al.}(2015){Hirota}, {Kim}, {Kurono}, \&
  {Honma}}]{Hirota2015}
{Hirota}, T., {Kim}, M.~K., {Kurono}, Y., \& {Honma}, M. 2015, \apj, 801, 82

\bibitem[{{Hirota} {et~al.}(2016){Hirota}, {Machida}, {Matsushita}, {Motogi},
  {Matsumoto}, {Kim}, {Burns}, \& {Honma}}]{Hirota2016b}
{Hirota}, T., {Machida}, M.~N., {Matsushita}, Y., {et~al.} 2016, \apj, 833, 238

\bibitem[{{Kim} {et~al.}(2008){Kim}, {Hirota}, {Honma}, {Kobayashi},
  {Bushimata}, {Choi}, {Imai}, {Iwadate}, {Jike}, {Kameno}, {Kameya},
  {Kamohara}, {Kan-Ya}, {Kawaguchi}, {Kuji}, {Kurayama}, {Manabe}, {Matsui},
  {Matsumoto}, {Miyaji}, {Nagayama}, {Nakagawa}, {Oh}, {Omodaka}, {Oyama},
  {Sakai}, {Sasao}, {Sato}, {Sato}, {Shibata}, {Tamura}, \&
  {Yamashita}}]{Kim2008}
{Kim}, M.~K., {Hirota}, T., {Honma}, M., {et~al.} 2008, \pasj, 60, 991

\bibitem[{{Kwan} \& {Scoville}(1976)}]{Kwan1976}
{Kwan}, J., \& {Scoville}, N. 1976, \apjl, 210, L39

\bibitem[{{Matthews} {et~al.}(2010){Matthews}, {Greenhill}, {Goddi},
  {Chandler}, {Humphreys}, \& {Kunz}}]{Matthews2010}
{Matthews}, L.~D., {Greenhill}, L.~J., {Goddi}, C., {et~al.} 2010, \apj, 708,
  80

\bibitem[{{Menten} \& {Reid}(1995)}]{Menten1995}
{Menten}, K.~M., \& {Reid}, M.~J. 1995, \apjl, 445, L157

\bibitem[{{Menten} {et~al.}(2007){Menten}, {Reid}, {Forbrich}, \&
  {Brunthaler}}]{Menten2007}
{Menten}, K.~M., {Reid}, M.~J., {Forbrich}, J., \& {Brunthaler}, A. 2007, \aap,
  474, 515

\bibitem[{{Peng} {et~al.}(2017){Peng}, {Qin}, {Schilke}, {S{\'a}nchez-Monge},
  {Wu}, {Liu}, {Li}, {M{\"o}ller}, {Liu}, {Feng}, {Liu}, {Luo}, {Zhang}, \&
  {Rong}}]{Peng2017}
{Peng}, Y., {Qin}, S.-L., {Schilke}, P., {et~al.} 2017, \apj, 837, 49

\bibitem[{{Pineau des Forets} {et~al.}(1993){Pineau des Forets}, {Roueff},
  {Schilke}, \& {Flower}}]{Pineau1993}
{Pineau des Forets}, G., {Roueff}, E., {Schilke}, P., \& {Flower}, D.~R. 1993,
  \mnras, 262, 915

\bibitem[{{Plambeck} \& {Wright}(2016)}]{Plambeck2016}
{Plambeck}, R.~L., \& {Wright}, M.~C.~H. 2016, \apj, 833, 219

\bibitem[{{Plambeck} {et~al.}(2003){Plambeck}, {Wright}, \&
  {Rao}}]{Plambeck2003}
{Plambeck}, R.~L., {Wright}, M.~C.~H., \& {Rao}, R. 2003, \apj, 594, 911

\bibitem[{{Plambeck} {et~al.}(2009){Plambeck}, {Wright}, {Friedel}, {Widicus
  Weaver}, {Bolatto}, {Pound}, {Woody}, {Lamb}, \& {Scott}}]{Plambeck2009}
{Plambeck}, R.~L., {Wright}, M.~C.~H., {Friedel}, D.~N., {et~al.} 2009, \apjl,
  704, L25

\bibitem[{{Plambeck} {et~al.}(2013){Plambeck}, {Bolatto}, {Carpenter},
  {Eisner}, {Lamb}, {Leitch}, {Marrone}, {Muchovej}, {P{\'e}rez}, {Pound},
  {Teuben}, {Volgenau}, {Woody}, {Wright}, \& {Zauderer}}]{Plambeck2013}
{Plambeck}, R.~L., {Bolatto}, A.~D., {Carpenter}, J.~M., {et~al.} 2013, \apj,
  765, 40

\bibitem[{{Snell} {et~al.}(1984){Snell}, {Scoville}, {Sanders}, \&
  {Erickson}}]{Snell1984}
{Snell}, R.~L., {Scoville}, N.~Z., {Sanders}, D.~B., \& {Erickson}, N.~R. 1984,
  \apj, 284, 176

\bibitem[{{Tsuboi} {et~al.}(1996){Tsuboi}, {Ohta}, {Kasuga}, {Murata}, \&
  {Handa}}]{Tsuboi1996}
{Tsuboi}, M., {Ohta}, E., {Kasuga}, T., {Murata}, Y., \& {Handa}, T. 1996,
  \apjl, 461, L107

\bibitem[{{Vaidya} \& {Goddi}(2013)}]{Vaidya2013}
{Vaidya}, B., \& {Goddi}, C. 2013, \mnras, 429, L50

\bibitem[{{Wright} {et~al.}(1995){Wright}, {Plambeck}, {Mundy}, \&
  {Looney}}]{Wright1995}
{Wright}, M.~C.~H., {Plambeck}, R.~L., {Mundy}, L.~G., \& {Looney}, L.~W. 1995,
  \apjl, 455, L185

\bibitem[{{Zapata} {et~al.}(2011){Zapata}, {Loinard}, {Schmid-Burgk},
  {Rodr{\'{\i}}guez}, {Ho}, \& {Patel}}]{Zapata2011a}
{Zapata}, L.~A., {Loinard}, L., {Schmid-Burgk}, J., {et~al.} 2011, \apjl, 726,
  L12

\bibitem[{{Zapata} {et~al.}(2009){Zapata}, {Schmid-Burgk}, {Ho},
  {Rodr{\'{\i}}guez}, \& {Menten}}]{Zapata2009}
{Zapata}, L.~A., {Schmid-Burgk}, J., {Ho}, P.~T.~P., {Rodr{\'{\i}}guez}, L.~F.,
  \& {Menten}, K.~M. 2009, \apjl, 704, L45

\end{thebibliography}
\clearpage
\end{document}